\documentclass[twocolumn,twoside,showkeys,showpacs,preprintnumbers,superscriptaddress,amsmath,amssymb,prc]{revtex4-2}

\usepackage{tikz}
\usepackage{float}
\usepackage[colorlinks=true,linkcolor=black,citecolor=blue,urlcolor=blue]{hyperref}

\usepackage{graphicx}% Include figure files
\usepackage{dcolumn}% Align table columns on decimal point
\usepackage{bm}% bold math
\usepackage[version=4]{mhchem}
\usepackage{subcaption}
\usepackage[colorlinks=true,linkcolor=black,citecolor=blue,urlcolor=blue,]{hyperref}
\usepackage{appendix}
\usepackage{ulem} 
\usepackage[justification=raggedright,singlelinecheck=false]{caption}
\usepackage{booktabs}

\usepackage{pstricks}
\usepackage{amsmath, amsthm, amssymb}
\usepackage{changes}
\definecolor{darkgreen}{RGB}{0,100,0}

\def\thu{Department of Physics, Tsinghua University, Beijing 100084, China}
\def\buaa{School of Physics, Beihang University, Beijing 100191, China}
\def\thuhep{Center for High Energy Physics, Tsinghua University, Beijing 100084, China}

\begin{document}
%\begin{CJK*}{UTF8}{gbsn}

%\preprint{APS/123-QED}

% \title{Precise Measurement of Short-Range Correlation in Nuclei through Bremsstrahlung $\gamma$-Rays in Nucleus-Nucleus Collisions}

%\title{Precise Measurement of Short-Range Correlations in Nuclei}%
%\title{Bremsstrahlung Gamma Ray as a Probe of Short-Range Correlations in Nuclei} 
\title{Calculation of Particle Pair Correlation Functions with Classical Trajectory Approximation}
% Force line breaks with \\
%\thanks{A footnote to the article title}%

\author{Sheng Xiao}
\email{xiaos20@mails.tsinghua.edu.cn}
\affiliation\thu
%\altaffiliation{Department of Physics, Tsinghua University, Beijing 100084, China}%Lines break automatically or can be forced with \\
\author{Yijie Wang}
\email{wyj25@buaa.edu.cn}
\affiliation\buaa 
\affiliation\thu 
%\affiliation{Department of Physics, Tsinghua University, Beijing 100084, China}%
%\collaboration{MUSO Collaboration}%\noaffiliation
\author{Zhigang Xiao}
%\email{xiaozg@tsinghua.edu.cn}
\affiliation\thu 
\affiliation\thuhep

\date{\today}

\begin{abstract}
 %Femtoscopy and Hanbury-Brown–Twiss (HBT) interferometry are significant to probe spatio-temporal source dynamics. 
% While traditional methods struggle with ultrashort-lived states, femtoscopy leverages momentum correlations to reconstruct source geometries and lifetimes. 

Femtoscopic interferometry is a powerful tool for probing the spatio-temporal evolution of emission sources in heavy ion reactions. A major challenge in the field is formulating a self-consistent description of the source function, final-state interactions between the particle pair, and interactions inherent to the source itself. To address this, we have developed a novel Monte Carlo model for calculating two-particle correlation functions in a classical trajectory approximation (CTA-\uppercase\expandafter{\romannumeral1}). The model incorporates self-consistently the emission source of thermal equilibrium and three-body final-state interactions.  Application of the model shows satisfactory fit to experimental data, revealing that the correlation function is highly sensitive to the source's spatio-temporal extent. In contrast, the temperature parameter governing the emitted particles' energy spectra has a negligible influence. Our approach offers the potential to extract the spatio-temporal information from the emission source, thereby advancing the applicability of femtoscopic interferometry in the Fermi energy domain.

%the incorporating temperature-dependent effects and Coulomb interactions within a classical mean-field approach. By integrating thermal emission profiles and Coulomb corrections, the model simulates particle emission dynamics under realistic conditions. Key innovations include embedding temperature dependence into the emission spectra and addressing long-range Coulomb interactions, significantly influencing intermediate-mass fragment (IMF) dynamics. Validated against experimental data, this approach offers enhanced precision in extracting source parameters, expanding the applicability of femtoscopic analyses in heavy ion reactions in Fermi energy domain.
\end{abstract}

\maketitle

\section{Introduction}

One of the primary objectives of studying heavy ion reactions (HIRs) in the Fermi energy domain is to gain insights into the equation of state of nuclear matter (nEOS) near the saturation point \cite{COLONNA2020103775,CIAMPI2025139815,PhysRevC.95.041602}. However, extracting the parameters of the nEOS is significantly complicated by the intricate dynamics inherent in HIRs. To address these challenges, a key priority is to decode the spatio-temporal information of the particle emission source formed during these reactions.

Intensity interferometry, known as femtoscopy, has been developed and widely applied in nuclear physics since Hanbury-Brown and Twiss (HBT) pioneered this method to measure the angular size of Sirius \cite{BROWN1, HanburyBrown:1956bqd}. As an indispensable tool, femtoscopy fundamentally works by measuring the correlation functions of particle pairs emitted with small relative momenta from the reaction zone \cite{Lisa:2005dd,Lisa:2008gf,PhysRevC.20.2267}. Femtoscopy achieves two main objectives. On one hand, it enables the inference of the geometry and lifetime of emitting sources \cite{Bauer199277,Gelbke199991,Kotte2005271,PhysRevC.93.024905}, as well as the neutron distribution profile as recently proposed \cite{LI2025139963,HJZ2025}. On the other hand, it probes the interaction strength between correlated particle pairs, such as nucleon-nucleon (including p-p and n-n), nucleon-hyperon and other like- and unlike-baryon pairs \cite{KOONIN197743, PhysRevC.57.1428, PhysRevLett.134.222301,STAR:2015kha}. For a comprehensive review, one can refers to \cite{Verde:2006dh}.

Careful treatment is required when calculating the correlation functions in HIRs, given that the emission sources evolve dynamically. Both final-state interactions (FSI) between the correlated particle pair and the influence of the source's potential field distort the final momenta of the two correlated particles. Several models have been developed to calculate the correlation functions, addressing these complexities.

The CRAB (Correlation After Burner) program is one such model, developed to compute the correlation functions and extract key physical parameters, including particle source sizes (source radii), flow parameters (e.g., elliptic flow), and source expansion velocities \cite{Lisa:2005dd, ZhangJing-Bo_2001}. Under the assumption that the influence of the emission source's potential field on final state particle pairs can be neglected, the CRAB program generates correlation functions from the phase space of the emitting source. This phase space is derived from transport simulations or Monte Carlo sampling. The final state interaction between the particle pair is described by the potential, which is input in solving the Schrödinger equation to calculate the relative motion wave function. The correlation function is then obtained through integration over the phase space. By comparing these computational results with experimental data, key physical information such as the timescales and sequence of particle emission can be determined \cite{Wang:2021mrv, Wang:2024nmh, Ghetti:2003zz}.

Another pivotal framework for correlation function calculations is the Lednicky-Lyuboshits (LL) model \cite{Lednicky:1981su}. This model starts from the correlation function of point-like source expressed through Bethe-Salpeter amplitude. By considering only s-wave interactions, it applies the effective range approximation to calculate the scattering cross-section using given scattering length and effective range parameters. The point-like source correlation function is then integrated over the source using the Kopylov-Podgoretsky (KP) formula \cite{Lednicky:1981su}, allowing the correlation function to be computed analytically. By accounting for Bose or Fermi statistical effects and final-state interactions, the interaction parameters of particle pairs \cite{PhysRevLett.134.222301, STAR:2015kha, STAR:2018uho} and emission source distributions \cite{Xu:2024dnd, Anchishkin:2006qp} can be extracted through fitting the experimental correlation function. Although the LL model treats two-body scattering exactly, it neglects the influence of the residual nucleus due to the technical difficulty in solving three-body problem.

In HIRs within the Fermi energy domain, the MENEKA model \cite{Elmaani1992TrajectoryCF} is widely applied for correlation function analysis. Unlike the analytical approach of the LL model, MENEKA employs a classical trajectory-based method to compute correlation functions, explicitly treating the three-body dynamics of the correlated particle pair and the recoiling source \cite{Elmaani1992TrajectoryCF}. As a Monte Carlo simulation program, MENEKA operates under three key assumptions:
(i) Particles are emitted from the surface of an excited nuclear source, with initial directions following a distribution of orbital angular momenta;
(ii) Initial emission energies of particles match either experimental spectra or theoretical predictions;
(iii) The time delay between the successive emission of the two correlated particles follows an exponential decay law. In practice, MENEKA numerically simulates the trajectories of emitted particles using small time steps until the particles exit the interaction range. Such  classical trajectory approach  is particularly suited for capturing the dynamic interplay between particle emission and source recoil in Fermi-energy heavy ion reactions, where quantum effects are less dominant than classical dynamical processes.

 %{\color{blue} For spatial distribution, deblurring technique originated in optics applications has been deployed in heavy ion reactions, enabling to image the source function  with sensitivity to the reaction geometry as well as to the initial profile of the source.}

  The femtoscopic method has increasingly been applied to infer the spatial distribution and temporal evolution of the emission source. In terms of temporal evolution, the exponential decay function has been widely utilized with considerable success \cite{PhysRevLett.71.2863}.  However, for spatial distribution, due to the complexity of nuclear reactions, the source is not necessarily spherical.  For example, exotic geometries such as disks or rings have been explored in the context of femtoscopic approaches \cite{GLASMACHER1993265,PhysRevLett.78.2084}.  Deblurring technique originated in optics applications has been developed and deployed in heavy ion reactions, enabling to image the source function  with sensitivity to the reaction geometry as well as to the initial profile of the source \cite{PhysRevC.108.L051603,NZABAHIMANA2023138247,Nzabahimana:2025ivc,Nzabahimana:2025ivc1}. These approaches constitute an effective pathway to integrate femtoscopic correlations with transport models, thereby enabling the investigation of the spatial configuration of emission sources. The experimental inference of anisotropic sources has been reported by the HADES collaboration, utilizing high-order flow harmonics generated in heavy ion collisions \cite{PhysRevLett.125.262301}. Therefore, to accurately describe the emission source in femoscopy applications, as well as to understand how the emitted particles experience the interactions from the source in certain distribution , it is essential to develop a unified framework that treats the source geometry and the corresponding potential field experienced by the emitted particles in a self-consistent manner. A notable example is the correlation function for pairs of intermediate mass fragments (IMFs). In such cases, three-body effects, including the influence of the source, are crucial, as the Coulomb interaction between the IMFs and the source cannot be neglected.

%{\color{blue} Thus, one realizes that }the geometry of the emission source and its potential, including both nuclear and Coulomb effects, are correlated and must be treated in a unified manner. A notable example is the calculation of the correlation function for pairs of intermediate mass fragments (IMFs). In such cases, three-body effects, including the influence of the source, are crucial, as the Coulomb interaction between the IMFs and the source cannot be neglected.

Motivated by the requirement of self-consistent treatments of the emission source in thermal equilibrium and three-body interactions, we present a Monte Carlo model based on the classical trajectory approximation (CTA-\uppercase\expandafter{\romannumeral1}). With two key improvements  $-$  refined self-consistent mean field calculations and optimized temperature parameters for the Gaussian-shaped emitting source $-$ this model has proven effective in calculating IMF correlation functions. It reliably captures the interplay between thermal emission, Coulomb repulsion, and three-body dynamics, making it a robust theoretical tool for interpreting experimental IMF correlation data in Fermi-energy heavy ion reactions.

In this paper, we describe the analytical derivation and application of the model. We begin with the thermal equilibrium source using kinetic theory and discuss the form of the mean field. Then, we determine our observables and apply the model to interpret experimental data. The paper is organized as follows: Section 2 presents the model construction, including initialization, source description, and the simulation of particle emission dynamics. Section 3 applies the model to interpret experimental data, while Section 4 concludes with a summary and outlook.

\section {Model Construction}

Our model follows a general workflow that proceeds as follows:

{\it Initial Conditions:} Define the parameters of the reaction system, including the beam energy, the charge and mass of both the projectile and the target, and the charge  and mass  of the particles to be emitted. From these parameters, the approximate size of the residual nucleus can be estimated.

{\it Mean Field Definition:} Specify the mean field of the residual nucleus as the emission source. This interaction should be represented by the central potential corresponding to the initial conditions.

{\it Thermal Equilibrium and Emission Spectra:} Input the temperature of the thermal equilibrium emission source. This temperature determines the energy spectra of the emitted particles.

{\it Particle Emission and Evolution:} Sample the emitted particles and calculate their evolutionary dynamics, considering both the interactions between particle pairs and the potential field of the source.

{\it Event Filtering:} Finally, assess whether the emitted particles meet the predefined detector criteria. If they pass, the event is recorded. This part relies on the specific detector setup defined by the user.

%\subsection{Initial Conditions}
%$\quad\quad$ We first choose which particle to be emitted. These would decide whose correlation function will be calculated. By inputting the information about the collision system including the beam energy, the charge and mass of the projectile nuclei and the target nuclei, we can determine the charge and mass of the residual nucleus which would make an impact of the central force potential.

\subsection{Thermal Equilibrium Treatment}
We begin by defining a source as a ``region of homogeneity," where emission is described by a Poisson process. To uphold the property of stationary independent increments, it is vital that the emission source remains in thermal equilibrium. It should also be noted that if the emission rate becomes significantly large, the Poisson process will transition into a state without thermal equilibrium. Although this condition necessitates further discussion, it is important as it encompasses  a type of explosive source. As the first step, if not specifically clarified, we will reaffirm the assumption of an emission source in thermal equilibrium.

One writes the Hamiltonian $H$ which represents a particle in central force field by
\begin{equation}\label{Hamiltonian-1}
H=\frac{p_x^2+p_y^2+p_z^2}{2m}+V(r)
\end{equation}
where $r$ is the distance between the particle and the origin, $m$ is the mass of the particle. Since  the source is  thermalized, the single-particle momentum distribution function of the emitted particles takes the following Boltzmann form
\begin{equation}\label{distribution-1}
f_p(\vec{p})=(2\pi m k_{\rm B}T)^{-\frac{3}{2}}\exp{\left(-\frac{p_x^2+p_y^2+p_z^2}{2mk_{\rm B}T}\right)}
\end{equation}
where $k_{\rm B}$ is the Boltzmann constant, and $T$ is named the temperature of the source.  Although the high energy tail of the particle spectrum usually deviates from Boltzmann distribution, the deviation brings insignificant impact to the results, and is hence neglected here.  Similarly, one can assume that the spatial distribution function is $f_x(\vec{r})$, and the phase-space distribution function can be written as

\begin{equation}\label{distribution-2}
f(\vec{r},\vec{p})=f_x(\vec{r})f_p(\vec{p})
\end{equation}

And we know Liouville's theorem,
\begin{equation}\label{Liouville-1}
\frac{\partial f}{\partial t}+\{f,H\}=0
\end{equation}

here  the  Poisson bracket of $\{f,H\}$ reads
\begin{equation}\label{eqn-1}
\{f,H\}=\sum (\frac{\partial f}{\partial q_i}\frac{\partial H}{\partial p_i}-\frac{\partial H}{\partial q_i}\frac{\partial f}{\partial p_i})
\end{equation}
where the summation runs over $q_i,p_i$, which represent  generalized coordinates and momenta, respectively.  A time-independent solution means $\frac{\partial f}{\partial t}=0$. Now substituting (\ref{Hamiltonian-1}), (\ref{distribution-1}), (\ref{distribution-2}) into (\ref{Liouville-1}), and setting $\frac{\partial f}{\partial t}=0$, one obtains

\begin{equation}\label{Liouville-2}
f_p(\vec{p})(\nabla f_x(\vec{r}))\cdot\frac{\vec{p}}{m}-f_p(\vec{p})f_x(\vec{r})(-\frac{\vec{p}}{mk_{\rm B}T})\cdot \nabla V(r)=0
\end{equation}
This equation holds for arbitrary $\vec{p}$, i.e.
\begin{equation}\label{Liouville-3}
\nabla f_x(\vec{r})+\frac{1}{k_{\rm B}T}f_x(\vec{r})\nabla V(r)=0
\end{equation}

The equation (\ref{Liouville-3}) indicates that the initial position of the emitted particle is linked to the central mean field $V(r)$, which is constrained by thermal equilibrium. If one took naively  $V(r)$ as an isotropic pure Coulomb potential, one could derive the distribution function  $f_x(\vec{r})$
 as $f_x=c \exp{\left(-\frac{\alpha}{k_{\rm B}T}\frac{1}{r}\right)}$, where $c$ is a constant. However, this solution cannot be normalized, as the pure Coulomb potential will lead to a catastrophic disintegration of the system. Clearly, the mean field provided by the emission source cannot be modeled solely by a Coulomb potential. A short-range nuclear potential associated with the source must  be considered.

 In order to obtain a reasonable form of the attractive mean field, we start from a commonly used Gaussian  source   $f_x$, i.e.

\begin{equation}\label{Gaussian-1}
f_x(\vec{r})=(2\pi\sigma_{\rm R}^2)^{-\frac{3}{2}}\exp{\left(-\frac{x^2+y^2+z^2}{2\sigma_{\rm R}^2}\right)}
\end{equation}

where $\sigma_{\rm R}$ is a parameter of the Gaussian source, which is usually regarded as the source  size parameter. By substituting (\ref{Gaussian-1}) into (\ref{Liouville-3}), one obtains

\begin{equation}\label{Gaussian-2}
-\frac{\vec{r}}{\sigma_{\rm R}^2} f_x(\vec{r})+\frac{1}{k_{\rm B} T}f_x(\vec{r})\nabla V(r)=0
\end{equation}

i.e.
\begin{equation}\label{Gaussian-3}
\nabla V(r)=\frac{k_{\rm B}T}{\sigma_R^2}\vec{r}
\end{equation}

The potential of Eq. (\ref{Gaussian-3}) represents a three-dimensional spherically symmetric harmonic oscillator.

It is shown that the source distribution $f(r)$ and the mean field $V(r)$ are interconnected. , i.e., the widely-used Gaussian source in thermal equilibrium yields a harmonic potential near $r\approx0$. 
%In approximation, using a Gaussian distribution to parameterize the source is still meaningful and applicable in the vast majority of cases.

Given that the nuclear force is a short-range interaction, for a general central attractive potential, one can rewrite the potential using a Taylor expansion around the equilibrium point 
\begin{equation}\label{V_per2}
V(\vec{r})=V(r)=-U_0+\frac{1}{2}\kappa r^2 + o(r^2)
\end{equation}
where $\kappa$ is a positive constant. By substituting (\ref{V_per2}) into (\ref{Gaussian-3}), one obtains
\begin{equation}\label{V_per3}
\kappa=\frac{k_{\rm B}T}{\sigma_R^2}=\frac{1}{3}\left(\left.\frac{\partial^2 V}{\partial x^2}\right|_0+\left.\frac{\partial^2 V}{\partial y^2}\right|_0+\left.\frac{\partial^2 V}{\partial z^2}\right|_0\right)
\end{equation}
It is seen that, a general potential can be constrained by the temperature $T$ and the source parameter $\sigma_{R}$, simply because the potential and the source are interconnected.

However, the simplified Gaussian source-harmonic potential scenario brings new problem.   Because obviously, particles can not escape from a pure harmonic potential. Thus,  the high-order term  $o(r^2)$  in Eq. (\ref{V_per2}) remains significant at large $r$, influencing the interactions in the final state and enabling the emission of particles.
The high-order term $o(r^2)$ is determined by the boundary condition at $r=\infty$. A natural option is that  the potential at $r=\infty$ reduces to Coulomb repulsion.

%Given that the nuclear force is a short-range interaction, perturbation theory can typically be applied. For a general central attractive potential, the potential can be rewritten using a Taylor expansion around the equilibrium point, up to second order. Without loss of generality, we set the equilibrium point at $r=0$. From the condition that $\frac{\partial V}{\partial x_i}=0$ at the equilibrium point, we can express the general form of the potential as:

%\begin{equation}\label{V_per1}
%V(\vec{r})=-U_0+\frac{1}{2}\sum_{ij} V_{ij}x_ix_j\quad i,j=1,2,3
%\end{equation}
%Here $U_0$ is the depth of the potential at $r=0$. The first derivative term vanishes because the potential is at equilibrium, and the second-order term describes the effective harmonic potential around $r=0$. Consider the isotropic condition, one further writes
%\begin{equation}\label{V_per2}
%V(\vec{r})=V(r)=-U_0+\frac{1}{2}\kappa r^2
%\end{equation}
%where $\kappa$ is a positive constant. By substituting (\ref{V_per2}) into (\ref{Gaussian-3}), one obtains
%\begin{equation}\label{V_per3}
%\kappa=\frac{k_{\rm B}T}{\sigma_R^2}=\frac{1}{3}\left(\left.\frac{\partial^2 V}{\partial x^2}\right|_0+\left.\frac{\partial^2 V}{\partial y^2}\right|_0+\left.\frac{\partial^2 V}{\partial z^2}\right|_0\right)
%\end{equation}

\subsection{Mean Field}

As mentioned above, a primary state of thermal equilibrium is assumed in the calculation. 
%Therefore, we require a harmonic oscillator-like potential, implying that the potential must be well-defined and stable. On the other hand, the Coulomb interaction is long-range, meaning that the mean field must decay as $1/r$  at large distances. 
Considering the central force potential of the residual nucleus, we can construct the mean field starting from a general formula, which can be divided into three parts.

(i) Electric term. Since a point charge is not physical at such small scales, we model the positive charge as having a specific density distribution. To remain general, we assume that the positive charge density follows a spherically symmetric Gaussian distribution. That is,

\begin{equation}\label{GC}
\rho_{+}(\vec{r})=Z_{\rm res}\mathfrak{e}(2\pi\sigma_{c}^2)^{-\frac{3}{2}}\exp{\left[-\frac{r^2}{2\sigma_{\rm c}^2}\right]}
\end{equation}
where $Z_{\rm res}$ is the residual charge number and $\mathfrak{e}$ is the unit charge. $\sigma_{\rm c}$ characterizes the spatial extent of the charge distribution  providing the Coulomb potential.  Now we solve the Poisson equation and times the emission particle charge. We have
\begin{equation}\label{V_GC}
V_{\rm c}(r)=\alpha\frac{1}{r}{\rm erf}(\frac{r}{\sqrt{2}\sigma_{\rm c}})
\end{equation}
where $\alpha=\frac{Z_{\rm res}Z_{1}\mathfrak{e}^2}{4\pi\epsilon_0}$ and ${\rm erf}(x)$ is the Gaussian error function.

(ii) Volume term. This part comes from the effective nuclear force, and the volume potential is always set to be Wood-Saxon form.
\begin{equation}\label{V_v0}
V_{\rm v}(r)=\frac{V_0}{1+\exp{\left(\frac{r-r_0}{d}\right)}}
\end{equation}
Let $\beta=e^{-r_0/d}$, one obtains
\begin{equation}\label{V_v1}
V_{\rm v}(r)=\frac{V_0}{1+\beta \exp{(r/d)}}
\end{equation}
where $V_0,r_0$ and $d$ are three parameters. $V_0$ is the depth of the potential trap. $r_0$ and $d$ are the effective radius and the surface diffusion coefficient, respectively. Clearly, since $r_0$  and $d$ are  positive, the inequality  $0<\beta<1$ is always satisfied.

(iii) Surface term. Just like providing surface absorption in optical model, this part always takes the form of the derivative of the Wood-Saxon function.
\begin{equation}\label{V_v2}
V_{\rm s}(r)=\frac{S_0 \exp{(r/d)}}{\left[1+\beta \exp{(r/d)}\right]^2}
\end{equation}\par
Now, the mean field would be represented as,
\begin{equation}\label{V_GCMF}
V_{\rm MF}=V_{\rm c}+V_{\rm v}+V_{\rm s}
\end{equation}
where $r$ is the distance between the particle and the origin point. 

Here we have 6 microscopic parameters,  $\sigma_{\rm c}$, $V_0$, $S_0$, $d$, $\beta$ and $U_0$  for the source potential. They would be determined as follows. $(1)$ While $r\rightarrow\infty$, $V_{\rm MF}$ must behave like a purely Coulomb potential. This leads directly to $\alpha=z_1Z_{\rm res}\mathfrak{e}^2/4\pi\epsilon_0$ where $z_1$ is the charge number of the emitted particle. $(2)$ While $r\rightarrow 0$, $V_{\rm MF}$ must behave like a harmonic oscillator. Here, one can construct the potential by the Taylor expansion to (\ref{V_GC}), (\ref{V_v1}) and (\ref{V_v2}).
\begin{equation}\label{V_GC_exp}
V_{\rm c}(r)=\frac{\alpha}{\sigma_{\rm c}}\sqrt{\frac{2}{\pi}}(1-\frac{1}{6\sigma_{\rm c}^{2}}r^2)+O(r^3)
\end{equation}
\begin{equation}\label{V_v1_exp}
V_{\rm v}(r)=\frac{V_0}{1+\beta}(1-\frac{\beta}{(1+\beta)d}r-\frac{\beta(1-\beta)}{2(1+\beta)^2d^2}r^2)+O(r^3)
\end{equation}
\begin{equation}\label{V_v2_exp}
V_{\rm s}(r)=\frac{S_0}{(1+\beta)^2}(1+\frac{1-\beta}{(1+\beta)d}r+\frac{1-4\beta+\beta^2}{2(1+\beta)^2d^2}r^2)+O(r^3)
\end{equation}
In order to treat the  Gaussian source in thermal equilibrium self-consistently, we set the $r$ term to be 0, and set the coefficient of the term $r^2$ to be $\frac{k_{\rm B}T}{\sigma_{\rm R}^2}$. By combining (\ref{Gaussian-3}), (\ref{V_GCMF}), (\ref{V_GC_exp}), (\ref{V_v1_exp}), (\ref{V_v2_exp}), we derive the constraint as follows.
\begin{equation}\label{constrains_GC1}
\frac{\alpha}{\sigma_{\rm c}}\sqrt{\frac{2}{\pi}}+\frac{V_0}{1+\beta}+\frac{s}{(1+\beta)^2}=-U_0
\end{equation}
\begin{equation}\label{constrains_GC2}
-\frac{V_0}{1+\beta}\frac{\beta}{(1+\beta)d}+\frac{S_0}{(1+\beta)^2}\frac{1-\beta}{(1+\beta)d}=0
\end{equation}
\begin{equation}\label{constrains_GC3}
\begin{aligned}
-\frac{\alpha}{6\sigma_{\rm c}^3}\sqrt{\frac{2}{\pi}}-\frac{V_0}{1+\beta}\frac{\beta(1-\beta)}{2(1+\beta)^2d^2}\\
+\frac{S_0}{(1+\beta)^2}\frac{1-4\beta+\beta^2}{2(1+\beta)^2d^2}=\frac{k_{\rm B}T}{2\sigma_{\rm R}^2}
\end{aligned}
\end{equation}
Here we define that
\begin{equation}\label{ParTran1}
\sigma_{\rm c}=\gamma_{\rm c}\sigma_{\rm R}
\end{equation}
\begin{equation}\label{ParTran2}
d=\gamma_{\rm d}\sigma_{\rm R}
\end{equation}
\begin{equation}\label{ParTran3}
U_{\rm c}=\frac{\alpha}{\sigma_{\rm c}}\sqrt{\frac{2}{\pi}}
\end{equation}
and solve the constrain equations, one writes
\begin{equation}\label{constrains_sov1}
V_0=-(U_{\rm c}+U_0)(1+\beta)(1-\beta)
\end{equation}
\begin{equation}\label{constrains_sov2}
S_0=-(U_{\rm c}+U_0)\beta(1+\beta)^2
\end{equation}
\begin{equation}\label{constrains_sov3}
\frac{\gamma_{\rm d}^2 U_{\rm c}}{U_{\rm c}+U_0}\left(\frac{k_{\rm B}T}{2U_{\rm c}}+\frac{1}{6\gamma_{\rm c}^2}\right)=\frac{\beta^2}{(1+\beta)^2}
\end{equation}
If one chooses $\gamma_{\rm c}$ and $\gamma_{\rm d}$ to be free, there will be only 3 parameters. However, the constraint of $0\textless \beta \textless 1$ shall be satisfied, and it leads to the following inequality.
\begin{equation}\label{constrains_hedr}
\frac{\gamma_{\rm d}^2 U_{\rm c}}{U_{\rm c}+U_0}\left(\frac{k_{\rm B}T}{2U_{\rm c}}+\frac{1}{6\gamma_{\rm c}^2}\right)\textless \frac{1}{4}
\end{equation}

 Out of the 6 initial parameters with 3 derived constraints, namely Eq. (\ref{constrains_sov1}), Eq. (\ref{constrains_sov2}) and Eq. (\ref{constrains_sov3}),  three free parameters are left. One can choose $\gamma_{\rm c}$, $\gamma_{\rm d}$ and $U_0$  as the free parameters for the potential, while $T$, $\sigma_{\rm R}$  are taken as input quantities, which define the characteristics of the thermal equilibrium emitting source.

Up to now, we have successfully defined the emitting source as a thermal equilibrium fireball with temperature $k_{\rm{B}}T$  and Gaussian source size $\sigma_{\rm R}$, where the charge distribution follows a Gaussian form with standard deviation $\sigma_{\rm{c}}$. The core, providing the attractive nuclear interaction, is governed by a potential trap characterized by $U_0$, and the surface diffusion is described by the coefficient $d$.

\begin{figure}[hbtp]
    \centering
    \includegraphics[width=0.48\textwidth]{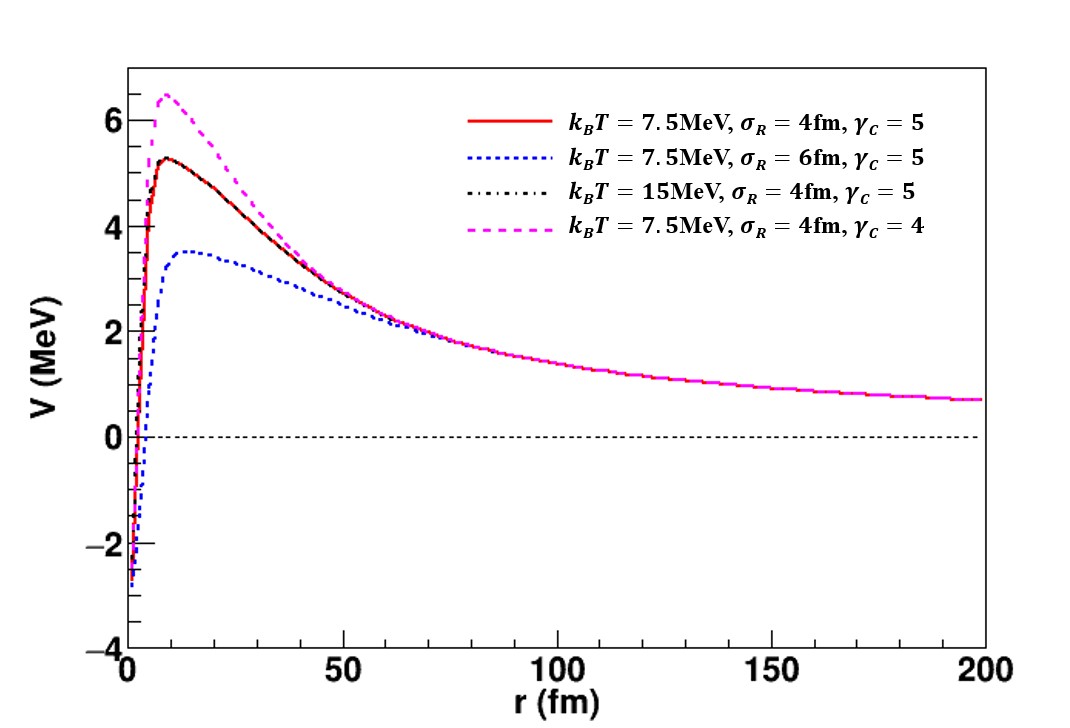}
    \caption{The mean field  experienced by a triton  at different settings of $k_{\rm B}T$, $\sigma_{\rm R}$ and $\gamma_{\rm c}$. Here  $\gamma_{\rm{d}}=0.3$ and $U_0=3$ MeV are fixed, with the reaction system is $^{86}$Kr+$^{208}$Pb.}
    \label{mfield}
\end{figure}

  Fig. \ref{mfield} presents a set of the potential of different parameters. It is shown that the temperature $k_{\rm B}T$ causes almost no difference, which is consistent with the picture that the temperature shall not make effect to the potential. The Gaussian source size $\sigma_{\rm R}$ takes effect both to the position and the height of the peak, while the $\gamma_{\rm c}$ influence only  the height.  

\subsection{Dynamics and Correlation}

In this subsection, we solve the dynamic evolution and derive the observable $-$ the correlation function.  

%The initial state of the motion is constructed by random sampling of the particle pair from  the emitting source. After the initial state is set,  the dynamics of the correlated particle pair can be simulated. This process is  treated by classic mechanics. Namely, the  trajectories of emitted particles are simulated through time series, with the interaction between the emitted particles and the source being taken into account.  Here we define each time step as $\Delta t$ time interval. If the pair of particles with mass $(m_1,m_2)$ follows the trajectory $({(\vec{x}_1(t),\vec{p}_1(t)),(\vec{x}_2(t),\vec{p}_2(t))})$, the state of the next time step would be calculated by the following steps. 

%First, calculate the test movement.
%At this stage, the emitting source has been fully defined. 

The initial state of the motion can now be constructed by random sampling based on the characteristics of the emitting source. The sequential action then involves the simulation of the dynamics of the correlated particle pair, which is described by classical mechanics. Specifically, the trajectories of the emitted particles are simulated over time through a time series, with each time step being defined by  $\Delta t$ as a time interval. The interactions between the emitted particles and the source are taken into account. 
For a pair of particles with masses $(m_1,m_2)$, their trajectories are described by the time-dependent positions and momenta, written as ${(\vec{x}_1(t),\vec{p}_1(t))}$ and ${(\vec{x}_2(t),\vec{p}_2(t))}$, respectively. The state of the system after one time step is calculated using the following steps.

First, the test movement is calculated as following

\begin{equation}\label{dynamics_x1}
\vec{x}'_i(t+\Delta t)=\vec{x}_i(t)+\frac{1}{m_i}\vec{p}_i(t)\Delta t-\frac{1}{2}\frac{\nabla_i V(\vec{x}_i;\vec{x}_j)}{m_i}\Delta t^2
\end{equation}
\begin{equation}\label{dynamics_p1}
\vec{p}'_i(t+\Delta t)=\vec{p}_i(t)-\nabla_i V(\vec{x}_i;\vec{x}_j))\Delta t
\end{equation}
where $(i,j)\in \{(1,2),(2,1)\}$.

Next, we assume that the average force during a time interval can be constructed by combining the force from the current state and the force from the test state. This results in the acceptable movement of the particle pair for the given time step.

\begin{equation}\label{dynamics_x1}
\begin{aligned}
\vec{x}_i(t+\Delta t)=\vec{x}_i(t)+\frac{1}{m_i}\vec{p}_i(t)\Delta t\\
-\frac{1}{2}\frac{(\gamma \nabla_i V(\vec{x}_i;\vec{x}_j)+(1-\gamma)\nabla_i' V(\vec{x}_i';\vec{x}_j'))}{m_i}\Delta t^2
\end{aligned}
\end{equation}
\begin{equation}\label{dynamics_p1}
\vec{p}_i(t+\Delta t)=\vec{p}_i(t)-(\gamma \nabla_i V(\vec{x}_i;\vec{x}_j)+(1-\gamma)\nabla_i' V(\vec{x}_i';\vec{x}_j'))\Delta t
\end{equation}
where $(i,j)\in \{(1,2),(2,1)\}$, and  $\gamma$ is a numerical parameter to balance the force  before (subscript `pre') and after (subscript `pos') a motion step, satisfying  $ F_{\rm cal}=\gamma F_{\rm pre}+(1-\gamma)F_{\rm pos}$. We choose $\gamma=0.5$ in calculation.

At this stage, the main frame of simulation model has been constructed. The final step is to incorporate the correlation between the particles. Since we are neglecting the effects of Bose-Einstein or Fermi-Dirac statistics, the correlation arises purely from the dynamical interactions between the particles.
The interaction between the pair of particles can be described by their position vectors $\vec{r}_1$ and $\vec{r}_2$, which represent the positions of the particles at a given time. The correlation between the particles is governed by the forces that result from their relative positions and the dynamics of their interaction.

\begin{equation}\label{V1}
V_1(\vec{r}_1;\vec{r}_2)=V_{\rm MF}(r_1)+V_{12}(\vec{r}_1-\vec{r}_2)
\end{equation}
\begin{equation}\label{V2}
V_2(\vec{r}_2;\vec{r}_1)=V_{\rm MF}(r_2)+V_{12}(\vec{r}_2-\vec{r}_1)
\end{equation}

%If considering only the Coulomb interaction between the two emitted particles {\color{red}( here also need clarification that the source intereaction is considered.)}, one writes
If the interaction between the emitted particle pair only contains the Coulomb interaction, one writes
\begin{equation}\label{V12}
V_{12}(\vec{r}_{12})=\frac{Z_1 Z_2 \mathfrak{e}^2}{4\pi\epsilon_0}\frac{1}{\vec{r}_{12}}
\end{equation}
where $Z_1$ and $Z_2$ are the charge numbers of the particle pair.

%\subsection{Correlation function}
While completing a series of run,  a set of final events can be accumulated.  The correlation function is then defined as
\begin{equation}\label{CF_1}
C(q)=1+R(q)=C_{12}\frac{\Sigma Y_{12}(\vec{p}_1,\vec{p}_2)}{\Sigma Y_1(\vec{p}_1)Y_2(\vec{p}_2)},
\end{equation}
where $\vec{p}_1,\vec{p}_2$ are the laboratory momentum, $Y_{12}$ is the coincidence yield, $Y_1,Y_2$ are the inclusive yields of single particle. Here $q=\mu |\vec{p}_1/m_1-\vec{p}_2/m_2|$ is the relative momentum of the correlation pair, where $\mu=m_1m_2/(m_1+m_2)$ is the reduced mass. The normalization constant $C_{12}$ is determined by the requirement of $C(q)=1$ at large relative momentum. In experiment, the correlation function is taken as the normalized ratio of the relative momentum distribution in the same event to that in the mixed event as
\begin{equation}\label{CF_exp}
C(q)=C_{12}\frac{Y_{\rm same}(q)}{Y_{\rm mix}(q)}
\end{equation}
where the subscript `same' and `mix' denote same event and mixed event, respectively.

\subsection{Parameterization}

%In addition to the parameters mentioned above, one still requires further 2 parameters regarding dynamics, as to be introduced in the next section.
For clearness, this subsection summarizes the parameterization scheme of the model. To control the flow of the calculation, the following parameter sets are required.

i) \textbf {Reaction system}. The parameter set defining the reaction system is written as
\begin{equation}\label{par_set_hic}
\mathcal{F}=\{(E_{\rm b},Z_{\rm p},A_{\rm p},Z_{\rm t},A_{\rm t},R_{\rm LMT})\}
\end{equation}
For $\forall b_{\rm f} \in \mathcal{F}$, $b_{\rm f}$ is a vector that defines collision condition, where $E_{\rm b}$ is the beam energy per nucleon, $Z$ and $A$ are the charge and the mass number of the projectile and the target, represented by the subscripts $p$ and $t$, respectively.  The parameter $R_{\rm LMT}$ is the ratio of linear momentum transfer.

In a simplified incomplete fusion picture of heavy ion reaction in Fermi energy domain, the emission source is associated with $R_{\rm LMT}$ , characterizing how much of the beam momentum is transferred to the residual system, or usually the target-like fragments. Simply considering the conservation laws of energy, momentum and mass number,  one can write  $R_{\rm LMT}$  as
\begin{equation}\label{par_LMR}
R_{\rm LMT}=\frac{Z_{\rm res}+Z_{1}+Z_{2}}{Z_{\rm p}+Z_{\rm t}}=\frac{A_{\rm res}+A_{1}+A_{2}}{A_{\rm p}+A_{\rm t}}
\end{equation}
$Z_{\rm res}$ and $A_{\rm res}$ are the charge and mass number of the residual nucleus. Meanwhile, the connection between the residual frame and the laboratory frame is a Galilean transformation with velocity $v_{\rm res}$.
\begin{equation}\label{v_res}
v_{\rm res}=\frac{\sqrt{2A_{\rm p}m_u E_{\rm b}}}{(A_{\rm p}+A_{\rm t})m_u}
\end{equation}
where  $m_u$ is the average mass per nucleon.

ii) \textbf{ Emission  source}. The parameter set defining the emission source is written as 
\begin{equation}\label{par_set_es}
\mathcal{P}=\{(k_{\rm B}T,\sigma_{\rm R},\gamma_{\rm c},\gamma_{\rm d},U_0,\tau)\}
\end{equation}
For $\forall b_{\rm p} \in \mathcal{P}$, $b_{\rm p}$ is a vector that defines the self-consistent emission source, where $k_{\rm B}T$ is the characteristic temperature, $\sigma_{\rm R}$ is the source size of the Gaussian source. Here the parameter $\tau$ is introduced to represent the reciprocal of the Poisson rate. Equivalently, the time distribution of the particle emission follows an exponential descend $p(t)\propto \exp(-t/\tau)$.
%{\color{red} Here one shall introduce a parameter of time scale}

iii)  \textbf{Emitted particle pair}. The parameter set defining the emitted particle pairs is written as
\begin{equation}\label{par_set_pp}
\mathcal{U}=\{(Z_1,Z_2,A_1,A_2,m_1,m_2)\}
\end{equation}
For $\forall b_{\rm u} \in \mathcal{U}$, $b_{\rm u}$ is a vector that defines the simulated particles, where $Z_i$, $A_i$ and $m_i$ ($i=1,2$) refer to the charge, mass number and the mass of the emitted particle $i$.

iv)  \textbf{Dynamic evolution}. The parameter set controlling the motion of the particle pair in the field of the source is written as
\begin{equation}\label{par_set_ct}
\mathcal{C}=\{(\Delta t,\gamma,t_{\rm max},r_{\rm max})\}
\end{equation}
For $\forall b_{\rm c} \in \mathcal{C}$, $b_{\rm c}$ is a vector that controls the accuracy of the simulation. The end point of the simulation procedure is controlled by $t_{\rm max}$ and $r_{\rm max}$.

v)  \textbf{Experimental filtering}. Optionally, the parameter set defining the detector acceptance is expressed as
\begin{equation}\label{det_setup}
\mathcal{D}=\{Detector~Setup\} 
\end{equation}

$\mathcal{D}$ is  related to the specific detector setup, importantly taking the geometric coverage and the momentum resolution into account. In order to achieve a precise comparison between the model prediction and the experimental data, all the accumulated events, defined by $\mathcal{E} \subset \mathcal{M}=\{(\vec{p}_1,\vec{p}_2)\}$, are filtered by the detector setup $\mathcal{D}$. The detector filtering procedure is necessarily implemented by the user. By writing the acceptable set as
\begin{equation}\label{par_set_geo}
\mathcal{G}=\{Acceptable~Events\},
\end{equation}
the filtering process is equal to perform an intersection operation. The set of final events detected is written as
\begin{equation}\label{accept_relation}
\mathcal{E_\mathcal{D}} = \mathcal{E} \cap \mathcal{G}
\end{equation}

Eventually, our simulation could be expressed as such a mapping, $f_s:\mathcal{F} \times \mathcal{P} \times \mathcal{U} \times \mathcal{C} \times R\longrightarrow \mathcal{M}$. Here, $R$ refers to the random number. If we do the simulation repetitively with different random numbers, we will get a set of final events $\mathcal{E}$. By implementing the detector filtering procedure, one can finally get the set of final events detected $\mathcal{E}_D$.

\section{Result and Discussions}

Up to this point, the entire framework of the model has been illustrated. The correlation function between two particles originating from the source can now be calculated numerically. This framework can be applied to experimental data, both for pairs of intermediate mass fragments (IMFs) and pairs of light charged particles (LCPs).

\begin{figure}[hbtp]
    \centering
    \includegraphics[width=0.5\textwidth]{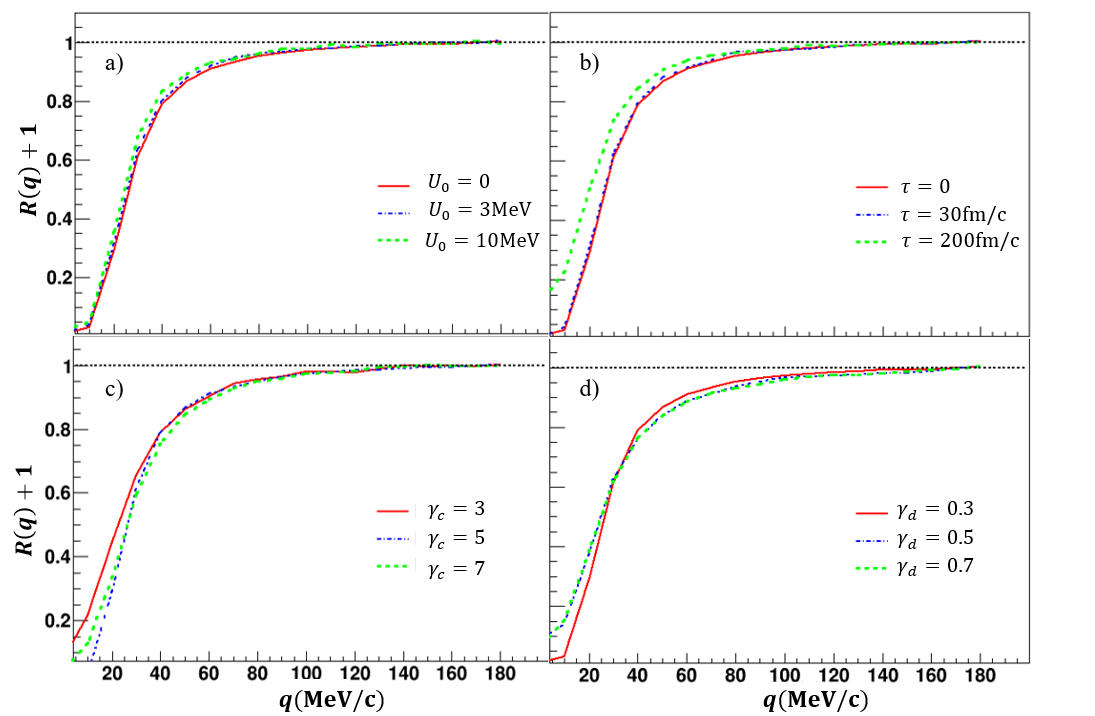}
    \caption{ Influence of the $\gamma_{\rm c}$, $\gamma_{\rm d}$, $U_0$ and $\tau $  on the correlation function of t-t pairs for the system by fixing other parameter at ($E_{\rm b}=25$ MeV/u, $Z_{\rm p}=36$,  $A_{\rm p}=86$,   $Z_{\rm t}=82$,  $A_{\rm t}=208$) $\in \mathcal{F}$. ($\Delta t=1$ fm/c, $\gamma=0.5$, $t_{\rm max}=12000$ fm/c,$r_{\rm max}=500$ fm) $\in \mathcal{C}$ and $\sigma_{R}=4$fm, $k_{\rm{B}}T=7.5$MeV. (a) Varying $U_0$ with fixed $\gamma_{\rm c}=5$, $\gamma_d=0.3$, $\tau =0$. b) Varying $\tau$ with fixed $\gamma_{\rm c}=5$, $\gamma_d=0.3$, $U_0=0$. c) Varying $\gamma_{\rm c}$ with fixed $\gamma_d=0.3$, $U_0=0$, $\tau =0$. d) Varying $\gamma_d$ with fixed $\gamma_{\rm c}=5$, $U_0=0$, $\tau =0$.}
    \label{impact_para}
\end{figure}

Before applying the model to the experimental correlation functions, we first check the impact of the main parameters on the correlation function. Fig. \ref{impact_para} presents the triton-triton correlations functions of different parameter sets. For each panel (a) to (d), only one parameter among $U_0$, $\gamma_c$,  $\gamma_d$ and $\tau$ is changed, while the others are fixed.  Let's recall that the parameter $U_0$ means the potential well depth of the residual nucleus. $\gamma_c$ means the ratio of the positive charge distribution size to the emission source size. $\gamma_d$ is the ratio of the surface diffusion coefficient of the residual nucleus to the emission source size. $\tau$ is the characteristic time interval of the emission which is equal to the reciprocal of the Poisson rate. It is shown that by varying these four parameters within a reasonable range, the correlation function exhibits less significant change as compared to the  influence of   emission source size $\sigma_R$ (see next). 

\subsection{Correlation functions of IMF-IMF pair}

We first apply the model to calculate the correlation functions of IMF pairs. The abundant IMFs emitted from Fermi energy HIRs carry crucial information about the violent reaction dynamics at the early stages. For instance, the correlation function of IMFs provides insights into the IMF emission timescale, which depends on the isospin reaction systems \cite{XIAO2006436}, as well as the space-time evolution of the colliding system \cite{PhysRevC.45.387,INDRA:2007tsb}. The experimental data is taken from approximate central reactions of Ar+Au  at 35 MeV/u beam energy, with the charged particles measured by the Miniball of Michigan State University \cite{PhysRevLett.67.14}. The correlation function is constructed by the IMFs detectors of the Ring 2 and Ring 3, situating at polar angles of $\theta_{\rm lab}=19.5^\circ$ and  $\theta_{\rm lab}=27^\circ$, respectively. For the details of the experiment, one can refer to \cite{PhysRevLett.67.14}.

\begin{figure}[hbtp]
    \centering
    \includegraphics[width=0.4\textwidth]{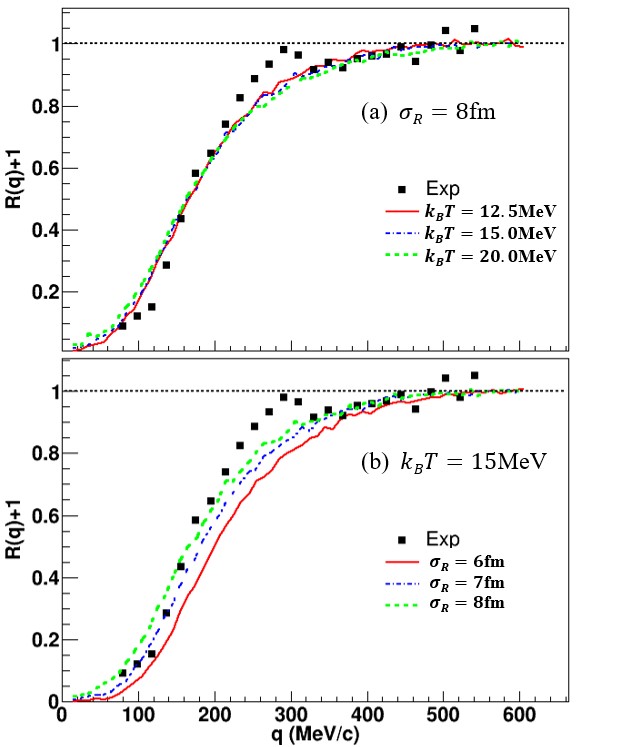}
    \caption{Correlation function of B-B pairs in comparison with the CTA-\uppercase\expandafter{\romannumeral1} model predictions for the $^{36}$Ar + $^{197}$Au reaction at $E/A=$35MeV/u .  (a) Varying $k_BT$ with fixed $\sigma_R=8$ fm. (b) Varying $\sigma_R$ with fixed $k_BT=15$ MeV. Data points are read off from \cite{PhysRevLett.67.14}, the statistic uncertainty is comparable to the symbol size}.
    \label{cf-bb}
\end{figure}

 For the interaction between the two correlated IMFs, it is reasonable to consider only the long-range Coulomb interaction. The mean field of the  emitting source, containing both Coulomb and nuclear potential, is taken into account. Figure \ref{cf-bb} presents the correlation function for Boron isotopes. Since the mass is not resolved, we set $A_1=A_2=10$ for the calculation.  For a rough comparison, here we skip the detector filtering procedure, since the efficiency loss is mostly canceled out when doing the ratio of the relative momentum spectrum in the same event to the mixed event, as shown in Eq. (\ref{CF_exp}). 

Fig. \ref{cf-bb} (a) compares the calculations with a fixed source size  $\sigma_{\rm R}= 8$ fm, while the temperature $k_{\rm B}T$ varies from 10 to 20 MeV. As expected, the correlation function shows negligible dependence on the temperature parameter. On the other hand, as shown in Fig. \ref{cf-bb} (b), where $k_{\rm B}T=15$ MeV is fixed, the source size  $\sigma_{\rm R}$  varies from 6 to 8 fm. Even though the variation of the source size is only 1 fm, it has a sensitive impact. The correlation becomes noticeably stronger when the source size decreases by 1 fm. With the parameters set at $k_{\rm B}T=15$ MeV and $\sigma_{\rm R}= 8$ fm, the experimental trend is well reproduced. These parameters are consistent with those extracted in \cite{PhysRevLett.67.14}.
It is worth mentioning that the small peak structure around  $q \approx 300$ MeV/c is not accounted for in this model, as it is unlikely to have  real physical correspondence.

\subsection{Correlation functions of LCP-LCP pair}

Finally, the model is applied to interpret the correlation function of LCP pair. Proton-proton (p-p) correlation function is not considered here, because it is not reliable to  treat the positive correlation peak due to the s-wave p-p resonance scattering in the classic context. Instead, we try the  triton-triton (t-t) and $^3$He-$^3$He pairs. The data are taken from the reactions 25 MeV/u $^{86}$Kr+$^{\rm nat}$Pb taken with the compact spectrometer for heavy ion experiment (CSHINE) \cite{GUAN2021165592}, which is installed at the final focal plane of the radioactive ion beam line at Lanzhou (RIBLL). 
%The beam of $^{86}$Kb was delivered by the heavy ion reach facility at Lanzhou bombarded on a natural lead target of 1 mg/cm$^2$. 
The charged particles were detected by 4 silicon strip detector (SSD) telescopes, each consisting of a single-sided SSD, a double-sided SSD and a $3\times 3$ CsI(Tl) array. The pixel size of each telescope is $4\times4$ mm$^2$, ensuring rather good position resolution. The energy resolution is better than 2\% \cite{Wang2021-ks}. A  track finding algorithm has been developed to identify the complicated firing pattern in the SSD telescopes \cite{Wei2025-eb,GUAN2022166461}. Three parallel plate avalanche counters (PPACs) were mounted to detect the fission fragments to reconstruct the event geometry. 
But for the correlation functions analysis here, no event geometry is selected because of the low statistics of four-body coincidence events. One can refer to \cite{GUAN2022166461,PhysRevC.107.L041601,Wang2025-vb} for the details of the experimental setup. 

\begin{figure}[hbtp]
    \centering
    \includegraphics[width=0.4\textwidth]{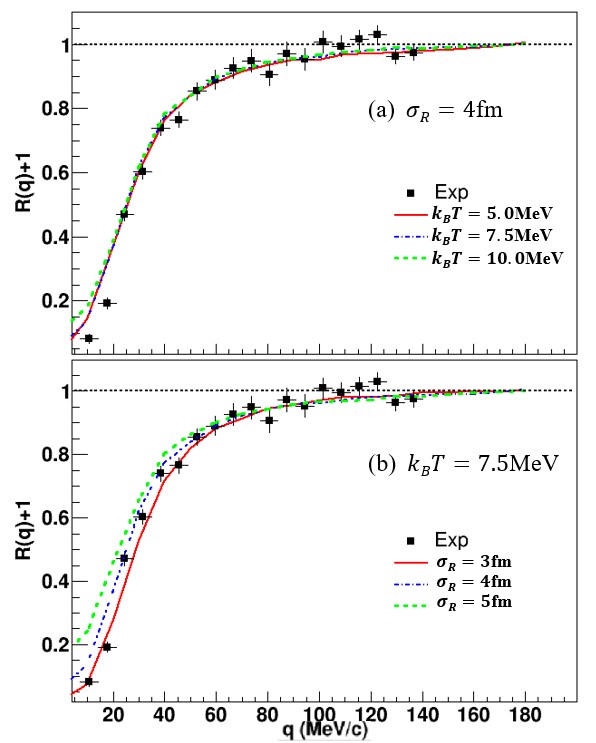}
    \caption{Correlation function of triton-triton pair in 25 MeV/u $^{86}$Kr+$^{\rm nat}$Pb reactions in comparison with the CTA-\uppercase\expandafter{\romannumeral1} model predictions. Statistic uncertainty of the data points are indicated. (a) Varying $k_BT$ with fixed $\sigma_R=4$ fm. (b) Varying $\sigma_R$ with fixed $k_BT=7.5$ MeV.}
    \label{cf-tt}
\end{figure}

Fig. \ref {cf-tt} presents the correlation functions of the t-t pair in comparison to the model calculations.  The mean $R_{\rm LMT}$ value is set to 0.8 in this analysis. As an example, the parameter settings are listed as following. ($E_{\rm b}=25$ MeV/u, $Z_{\rm p}=36$,  $A_{\rm p}=86$,   $Z_{\rm t}=82$,  $A_{\rm t}=208$) $\in \mathcal{F}$. ($\Delta t=1$ fm/c, $\gamma=0.5$, $t_{\rm max}=12000$ fm/c,$r_{\rm max}=500$ fm) $\in \mathcal{C}$. $\gamma_{\rm c}=5$, $\gamma_d=0.3$, $U_0=0$. Panel (a) and (b) present the calculations by varying $k_{\rm B}T$ and $\sigma_{\rm R}$, respectively. In panel (a), the source size parameter is fixed at $\sigma_{\rm R}=4$ fm. Again, the parameter $k_{\rm B}T$ shows much weak impact on the correlation function. In panel (b) where the  $k_{\rm B}T=7.5$ MeV is fixed, the correlation function exhibits sensitive dependence on the source size parameter $\sigma_{\rm R}$, in accordance with the picture that the correlation function can be used to probe the spatio-temporal size of the source. We note here that the temperature parameter for tritons is much lower in comparison with that of Boron in Fig. \ref{cf-bb}, because, in a qualitative picture, the emission of triton persists to relatively later stage due to its much lower Coulomb barrier.

%$\quad\quad$Here we have done some simulation with the parameter $(k_BT=62.5MeV, \sigma_R, %\beta=0.8) \in \mathcal{P}$, %$(E_{NN}=25MeV/u,Z_{projetile}=36,A_{projectile}=86,Z_{targrt}=82,A_{targrt}=208,LMR=0.9)\in %\mathcal{F}$, $(\Delta t=1fm/c,\gamma=0.5,t_{max}=12000fm/c,r_{max}=500fm)\in \mathcal{C}$. By %choosing a different $u\in \mathcal{U}$, we get the simulation result of different particles. %And we also calculate different correlation functions by varying %$\sigma_R=7.5fm,12.5fm,17.5fm$.

%$\quad\quad$Here we have done some simulation which could be compared with the CSHINE experimental data. We set $(E_{NN}=25MeV/u,Z_{projetile}=36,A_{projectile}=86,Z_{targrt}=82,A_{targrt}=208,LMR=0.8)\in \mathcal{F}$, $(\Delta t=1fm/c,\gamma=0.5,t_{max}=12000fm/c,r_{max}=500fm)\in \mathcal{C}$ and let $\gamma_C=5,\gamma_d=0.3$. By choosing different $k_BT$ and different $\sigma_R$, we fit the experiment data with $(k_BT=7.5MeV, \sigma_R=4fm)$ for $t-t$ pairs and $(k_BT=10.0MeV, \sigma_R=5fm)$ for $^{3}He-{^{3}He}$ pairs.

\begin{figure}[hbtp]
    \centering
    \includegraphics[width=0.4\textwidth]{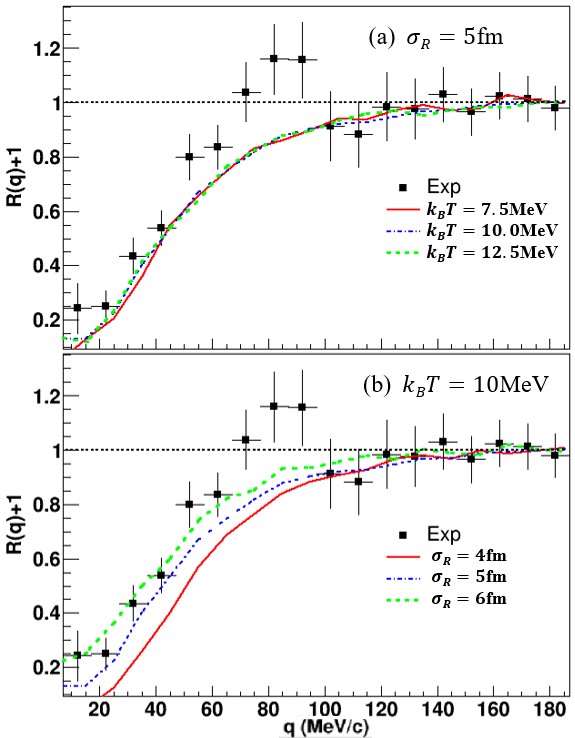}
    \caption{Correlation function of $\rm ^3He \text- ^3He$ pair in 25 MeV/u $^{86}$Kr+$^{\rm nat}$Pb reactions in comparison with the CTA-\uppercase\expandafter{\romannumeral1} model predictions. (a) Varying $k_BT$ with fixed $\sigma_R=5$ fm. (b) Varying $\sigma_R$ with fixed $k_BT=10$ MeV.}
    \label{cf-he3}
\end{figure}

Fig. \ref{cf-he3} shows the calculations of the $^3$He-$^3$He on top of the experimental data points. Because the reaction system is neutron-rich, the yield of $^3$He, and hence the correlation function of  $^3$He pair suffers from the low statistics. Nevertheless, the theoretic curves follow similar trend with varying $k_{\rm B}T$ and $\sigma_{\rm R}$. Namely, the variation of $k_{\rm B}T$ in a reasonable range brings less impact to the correlation, compared to the variation of source size parameter $\sigma_{\rm R}$. Despite the large fluctuation on the data points, the experimental trend   is in accordance with the calculation with  $k_{\rm B}T=10.0$ MeV and $\sigma_R=6$ fm, and the source-size dependence is more pronounced than that of the temperature.   

The model calculation unravels some subtle difference between t-t and $^3$He-$^3$He correlation functions. Comparing the model predictions in Fig. \ref{cf-he3} (b) and Fig. \ref{cf-tt} (b) for the same reaction system, it is seen that the change of the correlation function is more pronounced in $^3$He-$^3$He pair than  in t-t pair  with varying $\sigma_{\rm R}$ equally by 1 fm, because the anti-correlation arising from Coulomb interaction is much stronger in the former. Although different emission size is suggested between triton and $^3$He through the model-data comparison,  due to the large experimental uncertainty, it is not intended to extract the isospin effect of the source parameter here.  Nevertheless, it is expected that one can potentially probe the isospin effect of the particle emission from HIR process if reasonably high-quality  correlation function data are available for t-t and   $^3$He-$^3$He pairs.

 Worth mentioning, the current version of our model successfully reproduces the experimental correlation functions in various systems based solely on simple assumptions, namely the Poisson process and thermal equilibrium. This demonstrates the model's feasibility in describing the emission source in heavy ion reactions from a statistical perspective.
 In principle, the model can be extended to accommodate a variety of scenarios by incorporating more realistic descriptions of the experimental conditions. For instance:

    i) The spherical assumption can be relaxed, allowing the source to possess interesting geometries such as disks or rings. This transition leads to the source behaving as a rotating system while still maintaining the assumption of thermal equilibrium. This circumstance can be encountered in the fast fission events in heavy ion reactions \cite{Wang2025-vb,Diao2022}. In this case, analyzing the correlation functions (CFs) as a function of the direction of the relative momentum becomes interesting. 

    ii) If the emission rate is sufficiently high to drive the emission process from a Poisson process to a Gaussian process, the thermal equilibrium assumption will break down. In this non-equilibrium regime, the process can be treated within the relaxation time approximation, where the relaxation time is explicitly linked to the effective emission rate of the Gaussian process. Research on extending the current model to these situations is ongoing. 

\section{Conclusion}

In summary, we have developed a classical trajectory approximation model (CTA-\uppercase\expandafter{\romannumeral1},  here `I' means version 1.0) to calculate the correlation functions of particle pairs in heavy ion reactions within the Fermi energy domain. Assuming thermal equilibrium in particle emission, the model treats  self-consistently the effect of the residual nucleus and the three-body (the source and the particle pair) final-state interactions during the process. The model has been applied to interpret the experimental correlation functions of LCP-LCP and IMF-IMF pairs. Rather good consistency is observed between the model’s calculations and experimental data. It is demonstrated that the correlation function is not sensitive to the thermodynamic temperature but is sensitive to the Gaussian source size. While the thermodynamic temperature can typically be extracted from energy spectra, the CTA-\uppercase\expandafter{\romannumeral1} model provides a tool to constrain the Gaussian source size  and the Poisson rate in heavy ion reactions at Fermi energies. Prospectively, the framework of the model can be extended to  accommodate a variety of scenarios, including non-spherical source geometry and Gaussian process of high emission rate.

\textbf{{Code Availability Statement}} The source code and input file of  CTA-\uppercase\expandafter{\romannumeral1} are available upon request to S. Xiao.

\textbf{{Acknowledgement}} This work is supported by the National Natural Science Foundation of China under Grant Nos. 12335008 and 12205160, by the Ministry of Science and Technology under Grant No. 2022YFE0103400,  by the Center for High Performance Computing and Initiative Scientific Research Program in Tsinghua University.

\bibliographystyle{unsrt}
\bibliography{reference}
%\clearpage\end{CJK*}
\end{document}